\documentclass[prc,aps,floatfix,showpacs,twocolumn]{revtex4}
\usepackage{dcolumn}
\usepackage{bm}
\usepackage{color}
\usepackage{graphicx}
\usepackage{epsf}
\begin{document}
\hspace*{14.5cm}{\mbox{IFUP-TH}}
\title{Muon capture on deuteron and the neutron-neutron scattering length}
\author{L.E.\ Marcucci$^{\, {\rm a,b}}$, 
R.\ Machleidt$^{\, {\rm c}}$}
\affiliation{
$^{\,{\rm a}}$\mbox{Department of Physics, University of Pisa, 56127 Pisa, Italy}\\
$^{\,{\rm b}}$\mbox{INFN-Pisa, 56127 Pisa, Italy}\\
$^{\,{\rm c}}$\mbox{Department of Physics, University of Idaho, Moscow, Idaho 83844, USA}\\
}

\date{\today}

\begin{abstract}
\begin{description}
\item[Background:] 
We consider the muon capture reaction
$ \mu^- + \,^2{\rm H} \rightarrow \nu_\mu + n + n $, 
which presents a ``clean'' two-neutron ($nn$) system in the
final state. We study here its capture rate in the doublet 
hyperfine initial state ($\Gamma^D$).
The total capture rate for the muon capture 
$ \mu^- + \,^3{\rm He} \rightarrow \nu_\mu + \,^3{\rm H}$ ($\Gamma_0$)
is also analyzed, although, in this case,
the $nn$ system is not so ``clean'' anymore.
\item[Purpose:] 
We investigate whether $\Gamma^D$ (and $\Gamma_0$)
could be sensitive to the $nn$
$S$-wave scattering length ($a_{nn}$), and we 
check on the possibility to
extract $a_{nn}$ from an accurate measurement of $\Gamma^D$.
\item[Method:] 
The muon capture reactions are studied with nuclear 
potentials and charge-changing weak currents, derived within chiral effective
field theory. The next-to-next-to-next-to leading order (N3LO) chiral
potential  with cutoff parameter $\Lambda=500$ MeV is used,
but the low-energy constant (LEC) determining $a_{nn}$ 
is varied so as to obtain
$a_{nn}=-18.95$ fm, $-16.0$ fm, $-22.0$ fm, and $+18.22$ fm.
The first value is the present empirical one, while the last 
one is chosen such as to lead to a di-neutron bound system 
with a binding
energy of 139 keV. The LEC's $c_D$ and $c_E$, present in the
three-nucleon potential and  axial-vector current ($c_D$), are constrained
to reproduce the $A=3$ binding energies and the triton
Gamow-Teller matrix element.
\item[Results:] 
The capture rate $\Gamma^D$ is found to be $399(3)$ s$^{-1}$
for $a_{nn}=-18.95$ and $-16.0$ fm; and $400(3)$ s$^{-1}$ for 
$a_{nn}=-22.0$ fm. However, in the 
case of $a_{nn}=+18.22$ fm, 
the result of $275(3)$ s$^{-1}$ ($135(3)$ s$^{-1}$) is obtained, when the 
di-neutron system in the final state is unbound (bound). 
The total capture rate $\Gamma_0$ for muon capture on $^3$He
is found to be 1494(15) s$^{-1}$, 
1491(16) s$^{-1}$, 1488(18) s$^{-1}$, and 1475(16) s$^{-1}$ for
$a_{nn}=-18.95$ fm, $-16.0$ fm, $-22.0$ fm, and $+18.22$ fm, respectively.
All the theoretical
uncertainties are due to the fitting procedure and 
radiative corrections.
\item[Conclusions:] 
Our results
seem to exclude the possibility of constraining a negative $a_{nn}$ 
with an uncertainty of less than $\sim \pm 3$ fm through an 
accurate determination
of the muon capture rates, but the uncertainty on the present empirical value
will not complicate the interpretation of the (forth-coming) 
experimental results for $\Gamma^D$.
Finally, a comparison with the
already available experimental data discourages the 
possibility of a bound di-neutron state (positive $a_{nn}$). 
\end{description}
\end{abstract}


\maketitle
\section{Introduction}
\label{sec:intro}
Muon capture reactions on light nuclei,
in particular the 
$ \mu^- + \,^2{\rm H} \rightarrow \nu_\mu + n + n $ ($\mu$--2)
and
$ \mu^- + \,^3{\rm He} \rightarrow \nu_\mu + \,^3{\rm H}$ ($\mu$--3)
reactions, have recently attracted considerable attention,
both theoretically and 
experimentally~\cite{Mea01,Gor04,Kam10,Mar02,Mar11,Mar12a,Mar12b,Gol14}. 
One of the reasons for the interest in this issue is the on-going MuSun experiment
at the Paul Scherrer Institute (PSI), which is expected to reach a
precision of 1.5 \% in the measurement of the doublet
$\mu$--2 capture rate ($\Gamma^D$)~\cite{And07,Kam10}. 
In fact, the available
experimental data for $\Gamma^D$ are quite inaccurate: 
Wang {\it et al.} obtained
$\Gamma^D=365(96)$ s$^{-1}$~\cite{Wan65} more than forty years ago.
A few years later, Bertin {\it et al.} measured
$\Gamma^D=445(60)$ s$^{-1}$~\cite{Ber73}, while the 
measurements performed in the eighties yielded
$\Gamma^D=470(29)$ s$^{-1}$~\cite{Bar86}
and $\Gamma^D=409(40)$ s$^{-1}$~\cite{Car86}.
Note that all the experiments, except that of Ref.~\cite{Bar86},
used the neutron detection technique, i.e. detected a neutron
in the final state.
On the other hand,
for the $\mu$--3 total capture rate ($\Gamma_0$),
a very accurate measurement
is available~\cite{Ack98}, namely, $\Gamma_0=1496(4)$ s$^{-1}$.

Recent theoretical work on the $\mu$--2 and $\mu$--3 reactions
are summarized in Refs.~\cite{Mar11,Mar12a,Mar12b}.
In particular, the work of Ref.~\cite{Mar12b} represents 
the first attempt to apply to the considered processes 
a ``consistent''
chiral effective field theory ($\chi$EFT) approach. 
We briefly review it here: 
the considered two-nucleon ($NN$) potential is
that derived in $\chi$EFT up to
next-to-next-to-next-to leading order (N3LO) in the chiral
expansion by Entem and Machleidt~\cite{Ent03,Mac11}.
When applied to the $A=3$ systems, the $NN$ potential
is augmented by the three-nucleon
($NNN$) interaction derived at next-to-next-to leading order (N2LO),
in the local form of Ref.~\cite{Nav07}.
The charge-changing weak current
has been derived up to N3LO in Ref.~\cite{Par96}.  Its
polar-vector part is related, via the conserved-vector-current
constraint, to the (isovector) electromagnetic current, which
includes, apart from one- and two-pion-exchange terms,
two contact terms---one isoscalar and the other isovector---whose
strengths are parametrized by the low-energy constants (LEC's) 
$g_{4S}$ and $g_{4V}$.
The two-body axial-vector current includes terms of one-pion
range as well as a single contact current, whose strength is
parametrized by the LEC $d_R$.  The latter is related to the LEC
$c_D$, which, together with $c_E$, enters the N2LO $NNN$ 
potential~\cite{Gar-Gaz}. 
The cutoff $\Lambda$ of the momentum-cutoff function, 
needed to 
regularize potentials and currents before they can be
used in practical calculations, is
taken to be in the range (500--600) MeV.  
The LEC's $c_D$ (or $d_R$) and $c_E$
are determined with the following procedure: 
(i) the $^3$H and $^3$He wave functions are calculated with the
hyperspherical harmonics method (see Ref.~\cite{Kie08} for a review),
using the chiral potentials mentioned above.
The corresponding set of LEC's, $c_D$ and $c_E$, are determined by
fitting the $A=3$ experimental binding energies.
(ii) For each set of $c_D$ and $c_E$, the $^3$H and $^3$He wave
functions are used to calculate the Gamow-Teller (GT) matrix element
in tritium $\beta$-decay.
Comparison with the experimental value leads to a range of values for
$c_D$ for each cutoff parameter $\Lambda$, from which
the corresponding range for $c_E$ is determined.
Such a procedure has been widely used by now
in a variety of studies, like elastic few-nucleon scattering~\cite{Viv13},
electromagnetic structure of light nuclei~\cite{Pia13}, the
proton-proton weak capture~\cite{Mar13}, and the nuclear
matter equation of state up to third order 
in many-body perturbation theory~\cite{Cor14}.
Finally, after determining the LEC's $g_{4S}$ and
$g_{4V}$ by reproducing the $A=3$ magnetic
moments, it has been shown in 
Ref.~\cite{Mar12b} that the
consistent $\chi$EFT approach leads to predictions
(with an estimated theory uncertainty of about 1\%) for the rates of
muon capture on deuteron and $^3$He, that are in excellent
agreement with the experimental data.

Although extensively studied, 
a crucial aspect of the $\mu$--2 reaction
has not been enough investigated sofar: 
the $\mu$--2 reaction contains in the final
state a ``clean'' two-neutron ($nn$) 
system, and therefore 
the doublet capture rate $\Gamma^D$
could be sensitive to the $nn$
$S$-wave scattering length ($a_{nn}$). In the present work,
we check on this possibility, and investigate 
whether the $\mu$--2 reaction offers the possibility to
extract $a_{nn}$ from an accurate measurement
of $\Gamma^D$, as it will be available soon from the PSI 
experiment~\cite{And07,Kam10}. To this aim, we work in the same $\chi$EFT
framework as in Ref.~\cite{Mar12b}, but apply N3LO $NN$ potentials 
(with cutoff $\Lambda=500$ MeV~\cite{Ent03}) that
predict different values for $a_{nn}$, i.e., the empirical
value $a_{nn}=-18.95$ fm and two more values 
within a range of $\sim \pm 3$ fm from this empirical one.
Note that the empirical value 
has been obtained from pion
capture on the deuteron~\cite{Che07} and neutron-deuteron 
breakup experiments~\cite{Gon06}. We will consider also
a case for which $a_{nn}>0$, which leads to a shallow 
bound di-neutron state.
The reason behind this choice resides 
in the work of Ref.~\cite{Wit12}, where it was shown that a hypothetical
$^1S_0$ $nn$ bound state would affect the angular distributions of the 
neutron-deuteron elastic scattering and
deuteron breakup cross sections, although
a comparison to the available data 
for the total cross section and angular distributions could not 
decisively exclude the existence of such a bound state. 
The analysis was carried out based on the CD Bonn potential~\cite{Mac96},
where for $a_{nn}=+18.22$ fm a bound $nn$ state was found
with a binding energy $B_{nn}=0.144$ MeV. 

The paper is organized as follows: in Sec.~\ref{sec:calc} we present
the details of the calculation, and in Sec.~\ref{sec:res} we list and 
discuss the results. Our concluding remark are given in Sec.~\ref{sec:concl}.

\section{Calculation}
\label{sec:calc}

\begin{figure}[t]
\vspace*{0.3cm}
\includegraphics[width=3in,height=2in]{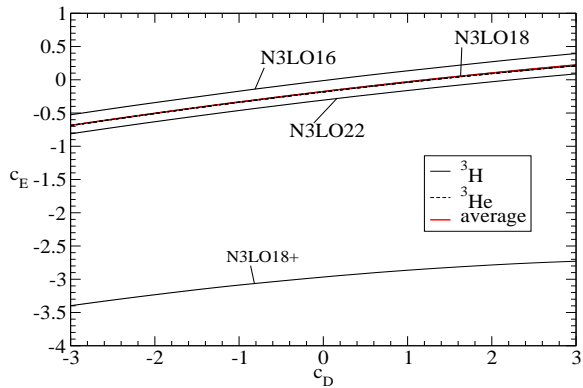}
\caption{(Color online) $c_D$-$c_E$ trajectories fitted to
reproduce the experimental $^3$H and $^3$He binding energies, for
the N3LO18, N3LO16, N3LO22, and N3LO18+ $NN$ potentials, 
augmented by the N2LO $NNN$ interaction 
model. 
Note that the curves that can be clearly distinguished in this figure are the 
$^3$H trajectories of the respective potentials. The corresponding $^3$He trajectories
cannot be distinguished and are essentially identical to the N3LO18 curve.
See text for further explanation.}
\label{fig:cdce_n3lo}
\end{figure}

\begin{figure}[t]
\vspace*{0.5cm}
\includegraphics[width=3in,height=2in]{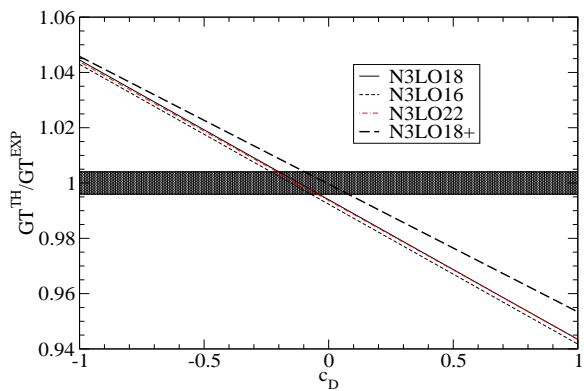}
\vspace*{0.1cm}
\caption{(Color online) 
The ratio GT$^{\rm TH}$/GT$^{\rm EXP}$ as function of the LEC $c_D$
for the N3LO18, N3LO16, N3LO22 and N3LO18+ $NN$ potentials and
the N2LO $NNN$ interaction model. The N3LO18 and N3LO22 lines are
essentially identical, and are very close to the N3LO16 one.}
\label{fig:gt}
\end{figure}

We summarize the various steps of our 
calculations. 
We consider the $NN$ potential at N3LO of Entem and Machleidt~\cite{Ent03},
with cutoff value fixed at $\Lambda=500$ MeV. The N3LO $NN$ potential
includes a charge-symmetry breaking 
contact term without
derivatives that contributes only in the $^1S_0$ state~\cite{Mac11}.
This contact is used to create different values for $a_{nn}$, which are,
in particular, 
 $-18.95$ fm, $-16.0$ fm and $-22.0$ fm.
We refer to these different versions of the $NN$ potential
as N3LO18, N3LO16, and N3LO22, respectively. The value $a_{nn}=-18.95$ fm 
corresponds to the empirical one.
Finally, we have also constructed a version of the N3LO potential,
which produces
$a_{nn}=+18.22$ fm (N3LO18+), leading to a two-neutron bound state,
with binding energy $B_{nn}=0.139$ MeV.
Then, for each given $NN$ potential, we add the N2LO $NNN$ interaction, 
and calculate the $^3$H and $^3$He binding energies as function of
the LEC's $c_D$ and $c_E$. 
The corresponding 
$c_D-c_E$ trajectories are given in 
Fig.~\ref{fig:cdce_n3lo}. 
Note that the trajectories which reproduce
the $^3$He binding energy for the various potentials are all on top of each other. Moreover, 
in the case of the N3LO18 potential, the $^3$H trajectory
is essentially the same as the $^3$He one.
However, for the other $NN$ potentials, the $^3$H trajectories differ from 
the corresponding $^3$He ones and from each other. 
This is particularly pronounced in the case of N3LO18+. 
For all cases, except N3LO18,
no average curve is displayed, and
all the $A=3$ wave functions
have been calculated using, for a given $c_D$, two different values of $c_E$,
one for $^3$H and one for $^3$He, i.e. allowing for 
charge-symmetry-breaking in the $NNN$ interaction.
Finally, using the
$\chi$EFT weak axial current of Ref.~\cite{Mar12b}, 
as discussed in Sec.~\ref{sec:intro}, the GT
matrix element of tritium $\beta$-decay (GT$^{\rm TH}$) is
determined. The ratio GT$^{\rm TH}$/GT$^{\rm EXP}$
is shown in Fig.~\ref{fig:gt}, for all $NN$ potentials.
The value GT$^{\rm EXP}=0.955 \pm 0.004$ has been used, 
as obtained in Ref.~\cite{Mar12b}.  The range of $c_D$ values for which
${\rm GT}^{\rm TH}={\rm GT}^{\rm EXP}$ within the experimental
error, and the corresponding ranges for $c_E$ are given in
Table~\ref{tab:cdce}. A few comments are in order: 
(i) in the N3LO18 case, the $^3$H and $^3$He values for $c_E$ are the 
same, since, as mentioned above, no charge-symmetry-breaking effect
is needed in the $NNN$ interaction (see Fig.~\ref{fig:cdce_n3lo}).
(ii) The $^3$He values
for $c_E$ are all close to each other. This reflects the fact that
the $np$ and $pp$ interactions are not affected 
by varying the LEC in the $NN$ potential to obtain different
$a_{nn}$ values. The small difference between the various $^3$He values
is due to the different range of $c_D$ as obtained by the GT 
fitting
procedure. This is again due to the different $nn$ interaction, which affects
the $^3$H wave function. (iii) The values for $c_E$ in the
$^3$H case are quite different between each other, especially
in the N3LO18+ case. Here we should remark that by using the N3LO18+
potential alone, i.e., without the $NNN$ interaction, the triton
and $^3$He binding energies are found to be 
9.935 MeV and 7.128 MeV, respectively, with an overbinding
in the case of the triton and an underbinding in the case of $^3$He. 
A large difference in the range for $c_D$ as well as a 
strong charge-symmetry-breaking effect is therefore natural.
(iv) The values for
$c_E(^3{\rm H})$ in the case of N3LO18+ are quite large,
and, according to the general trend of $\chi$EFT, should be considered
unnatural.

\begin{table*}[bthp]
\begin{center}
\caption{The LEC's $c_D$ and $c_E$ as obtained with the fitting procedure
explained in the text.}
\begin{tabular}{cccc}
\hline
$NN$ potential model  &  $c_D$ & $c_E(^3{\rm H})$ & $c_E(^3{\rm He})$\\
\hline
N3LO18  & $\{ -0.198 , -0.040 \}$ & $\{ -0.208 , -0.184 \}$ &$\{ -0.208 , -0.184 \}$ \\ 
N3LO16  & $\{ -0.231 , -0.072 \}$ & $\{ -0.047 , -0.023 \}$ &$\{ -0.218 , -0.194 \}$ \\ 
N3LO22  & $\{ -0.206 , -0.046 \}$ & $\{ -0.337 , -0.313 \}$ &$\{ -0.214 , -0.190 \}$ \\ 
N3LO18+  & $\{ -0.096 , +0.078 \}$ & $\{ -2.978 , -2.958 \}$ &$\{ -0.197 , -0.171 \}$ \\ 
\hline
\end{tabular}
\label{tab:cdce}
\end{center}
\end{table*}

The remaining LEC's $g_{4V}$ and $g_{4S}$ 
entering the $NN$ contact terms of the electromagnetic current 
have been fixed by reproducing the $A=3$ magnetic moments. Their values
are listed in Table~\ref{tab:LECs}. Notice that the values
of the LEC's in the N3LO18+ case are very different than in the other cases,
and, although of no relevance in the present study, the isoscalar LEC
$g_{4S}$ has even an opposite sign as compared to the N3LO18, N3LO16
and N3LO22 cases. Finally, 
the number in parentheses are the theoretical errors arising from 
numerics as explained in Ref.~\cite{Mar11}.

\begin{table*}[bthp]
\caption{The LEC's $g_{4S}$ and $g_{4V}$ associated with the
isoscalar and isovector $NN$ contact terms in the electromagnetic
current for the different $NN$ potentials considered here.
The number in parentheses are the theoretical errors, due to numerics.}
\begin{tabular}{cccc}
\hline
$NN$ potential model  &  $c_D$ & $g_{4S}$ & $g_{4V}$ \\
\hline
N3LO18 & $\{ -0.198 , -0.040 \}$ & $\{ 0.207(7) , 0.200(7) \}$ 
       & $\{ 0.765(4) , 0.771(4) \}$ \\
N3LO16 & $\{ -0.231 , -0.072 \}$ & $\{ 0.254(7) , 0.257(7) \}$ 
       & $\{ 0.801(4) , 0.804(4) \}$ \\
N3LO22 & $\{ -0.206 , -0.046 \}$ & $\{ 0.158(7) , 0.154(7) \}$ 
       & $\{ 0.747(4) , 0.745(4) \}$ \\
N3LO18+& $\{ -0.096 , +0.078 \}$ & $\{ -0.357(7) , -0.358(7) \}$ 
       & $\{ 0.463(4) , 0.471(4) \}$ \\
\hline
\end{tabular}
\label{tab:LECs}
\end{table*}

\section{Results}
\label{sec:res}

The results for the $\mu$--2 doublet capture rate $\Gamma^D$, also when only
the $^1S_0$ $nn$ partial wave is retained [$\Gamma_0(^1S_0)$], 
calculated with the different
$NN$ potential models, are listed in Table~\ref{tab:res}. The numbers
in parentheses are the theoretical uncertainties obtained by
summing, in a very conservative way, those arising from the 
LEC's fitting procedure and those present in the electroweak radiative
corrections~\cite{Cza07}. By inspection of the table, we can conclude
that the $\mu$--2 doublet capture rate is not sensitive to a
variation of $a_{nn}$ by $\sim \pm 3$ fm, as the change in $\Gamma^D$,
and $\Gamma^D(^1S_0)$ as well, is smaller than the theoretical
uncertainty of 1\% or less.  
On the other hand, a large difference is present for the N3LO18+ results,
as $\Gamma^D(^1S_0)$ is a factor of almost 2 smaller than in the other
cases. This reflects on $\Gamma^D$ as well, although the contributions
from the waves other than the $S$-wave remain unchanged, and this
reduces the difference between the N3LO18+ result and all the others
to a factor of $\sim 1.5$. Note that the
already available experimental data on $\Gamma^D$
obtained with the neutron detection technique,
$365(96)$ s$^{-1}$~\cite{Wan65}, 
$445(60)$ s$^{-1}$~\cite{Ber73}, 
and $409(40)$ s$^{-1}$~\cite{Car86}, 
although affected by
large uncertainties, seem to rule out the N3LO18+ case. 
For completeness we note
that in the case of a bound di-neutron $^1S_0$
state, $(nn)_b$, the reaction $\mu$--2 could go through the
channel 
$\mu^- + \,^2{\rm H} \rightarrow \nu_\mu + (nn)_b$
(subsequently denoted by $`\mu$--2b'). We have studied
the $\mu$--2b doublet capture rate and found
$\Gamma^D_b=135(3)$ s$^{-1}$.
By summing this value with the one listed in Table~\ref{tab:res},
we obtain $\Gamma^D=410(6)$ s$^{-1}$, where again,
in a very conservative way, we have linearly combined the theoretical
uncertainties. 
Notice, however, that this result is irrelevant in regard to the
experiments of Refs.~\cite{Wan65,Ber73,Car86}, because they all used
the neutron detection method,
which implies that they measured $\mu$--2 without a bound $nn$ state
contribution.

\begin{table}[bthp]
\caption{The $\mu$--2 doublet capture rate $\Gamma^D$ and the 
$\mu$--3 total capture rate $\Gamma_0$, in s$^{-1}$, calculated
using the $NN$ potential models
N3LO18, N3LO16, N3LO22 and N3LO18+, augmented, in the $\mu$--3 case, 
by the N2LO $NNN$ interaction. The values for $\Gamma^D$ when
only the $^1S_0$ $nn$ final state is retained are also listed.
The values in parentheses are the theoretical uncertainties.}
\begin{tabular}{cccc}
\hline
$NN$ potential model & $\Gamma^D(^1S_0)$ & $\Gamma^D$ & $\Gamma_0$ \\
\hline
N3LO18 & 254(2) & 399(3) & 1494(15) \\
N3LO16 & 254(2) & 399(3) & 1491(16) \\
N3LO22 & 255(2) & 400(3) & 1488(18) \\
N3LO18+& 130(2) & 275(3) & 1475(16) \\
\hline
\end{tabular}
\label{tab:res}
\end{table}

In Table~\ref{tab:res} we list also the results for the
the $\mu$--3 total capture rate $\Gamma_0$, although, in this case,
the $nn$ system is not ``clean'' anymore, as in the 
$\mu$--2 case. From these results, we can conclude
that the N3LO18+ case
is significantly smaller than all the others, but only 
slightly in disagreement with the accurate 
experimental datum of 1496(4) s$^{-1}$~\cite{Ack98}, due to our
theoretical uncertainty of about 1 \%.

\section{Summary and conclusions}
\label{sec:concl}

The muon capture reaction $\mu$--2 and $\mu$--3 have been studied
with nuclear 
potentials and charge-changing weak currents derived within $\chi$EFT.
The LEC present in the N3LO $NN$ potential
determining $a_{nn}$ is varied so as to obtain
values within a range of $\sim \pm 3$ fm around the empirical one.
A positive value for $a_{nn}$ has also been considered, such as
to lead to a di-neutron bound system with a binding
energy of 139 keV. 
Our results can be summarized as follows: 
no significant sensitivity to the $S$-wave $nn$ scattering length
is found, when this is changed within $\sim \pm 3$ fm from the 
present empirical value. The change in $a_{nn}$ affects
only the values for the LEC's $c_D$ and $c_E$, requiring a
charge-symmetry-breaking $NNN$ interaction, 
which has been taken into account in the present study.
The situation is quite different in the case of $a_{nn}=+18.22$ fm 
(case of $nn$ bound state),
since $\Gamma^D$ turns out to be a factor of about 1.5 smaller than in the 
previous cases, unless the $\mu$--2b reaction is included.
It should be mentioned also that
in the case of a positive $nn$ scattering length, 
the charge-symmetry-breaking effect in the $NNN$ interaction is
found to be very large, since the $^3$He nucleus  
is underbound, while the triton is
overbound. 

Therefore, we can conclude that 
the very accurate determination of $\Gamma^D$ 
by the MuSun collaboration at PSI will not be able to extract a 
more precise value for 
$a_{nn}$ in the case of a negative $a_{nn}$ value.
On the other hand, the uncertainty on the present empirical value
for $a_{nn}$ will not complicate the interpretation of the MuSun result,
from which it will be possible to obtain a clear extraction 
of the LEC $d_R$ (or $c_D$). Furthermore, the MuSun experiment
will most likely be able to confirm or exclude 
the existence of a bound di-neutron state,
if the experimental capture rate will have a sufficiently high accuracy
such that a value of 410(6) s$^{-1}$ can be ruled out.
However, in our opinion, 
should the existence of a bound di-neutron state 
be confirmed, then
our current very successful picture of muon
capture processes and, more general, of light nuclei 
would have to be severely revised. This is a similar
conclusion to what was obtained in Ref.~\cite{Pie03}, where the
possibility of a bound tetra-neutron system was investigated.

\section*{Acknowledgements}
We would like to thank P.\ Kammel for encouraging us to
carry out this study and for useful discussions. 
The work by R.M.\ was supported in part by the U.S.\ Department of Energy
under Grant No.~DE-FG02-03ER41270. Finally, we would 
like to acknowledge the assistance of the staff of the INFN-Pisa computer
center, where the calculations here presented were performed.

\end{document}